\documentclass[12pt]{iopart}
\usepackage{iopams}
\usepackage{epsfig}

\begin{document}

\title[The strange-quark chemical potential as an
experimentally accessible order parameter]{The strange-quark
chemical potential as an experimentally accessible "order
parameter" of the deconfinement phase transition for finite
baryon-density}

\author{Apostolos D. Panagiotou and Panayiotis G. Katsas\dag
}

\address{\dag\ University of Athens, Physics Department
Nuclear and Particle Physics Division, GR-15771 Athens, Hellas}

\ead{apanagio@phys.uoa.gr, pkatsas@phys.uoa.gr}

\begin{abstract}
We consider the change of the strange-quark chemical potential in
the phase diagram of nuclear matter, employing the Wilson loop and
scalar quark condensate order parameters, mass-scaled partition
functions and enforcing flavor conservation. Assuming the region
beyond the hadronic phase to be described by massive, correlated
and interacting quarks, in the spirit of lattice and effective QCD
calculations, we find the strange-quark chemical potential to
change sign: from positive in the hadronic phase - to zero upon
deconfinement - to negative in the partonic domain. We propose
this change in the sign of the strange-quark chemical potential to
be an experimentally accessible order parameter and a unique,
concise and well-defined indication of the quark-deconfinement
phase transition in nuclear matter.

\end{abstract}


\section{Introduction}

It is generally expected that ultra-relativistic nucleus-nucleus
collisions will provide the basis for strong interaction
thermodynamics, which will lead to new physics. Quantum
Chromodynamics for massless quarks free of dimensional scales
contain the intrinsic potential for the spontaneous generation of
two scales: one for the confinement force, coupling quarks to form
hadrons and one for the chiral force, binding the collective
excitations to Goldstone bosons. In thermodynamics, these two
scales lead to two possible consecutive phase transitions,
deconfinement and chiral symmetry restoration, characterized by
corresponding critical temperatures: $T_d$ and $T_{\chi}$, [1]. At
temperatures above $T_d$, hadrons dissolve into quarks and gluons,
whereas at $T_{\chi}$, chiral symmetry is fully restored and
quarks become almost massless (assuming that a thermal mass is
acquired for $T>T_d$), forming the Quark - Gluon Plasma (QGP). A
priori it is not evident that both non-perturbative transitions
have to take place at the same temperature. At finite net baryon
number density, $T_d<T_{\chi}$ would correspond to a regime of
unbound, massive, correlated and interacting, 'constituent-like'
quarks [2], as they appear in the additive quark model for
hadron-hadron and hadron-lepton interactions [3]. Thus, the
consecutive appearance of these two transitions, deconfinement and
chiral symmetry restoration, forms an intermediate region on the
phase diagram, the Deconfined Quark Matter (DQM), in-between the
Hadron Gas (HG) and the chiral QGP domains. Therefore, we define a
QCD phase diagram of strongly interacting matter at finite baryon
number density with three regions: HG - DQM - QGP [4-6].

The necessity for such an intermediate region with interacting
quarks and gluons is conjectured also from recent lattice
calculations [7], which show that $\varepsilon(2T_d) \sim
0.85\varepsilon_{SB}$ and $3P(2T_d) \sim 0.66\varepsilon_{SB}$,
where $\varepsilon_{SB}$ is the Stefan-Boltzmann ideal QGP value
for the energy density. In addition, the running coupling constant
has a value $\alpha_s(T \sim 300 MeV) \sim 0.3$, indicating
substantial strong interaction even at this high temperature.
These observations substantiate quantitatively the need for an
intermediate domain with $1 > \alpha_s > 0$, where massive and
correlated-interacting quarks are found. This domain does not have
a defined upper border, but goes asymptotically into the chirally
symmetric QGP region with increasing temperature and quark
chemical potential. At zero baryon density, $T_d = T_{\chi} =
T_0(\mu_q=0)$, and the DQM domain vanishes.

Such a 3-state phase diagramme could be described by the variation
of thermodynamic quantities from one region to the other. The task
is to establish a well-defined quantity, which changes concisely
(and measurably) as nuclear matter changes phases, thus indicating
deconfinement and / or chiral symmetry restoration. If this
quantity can be expressed in a functional form of other
thermodynamic parameters (for example the temperature) within an
Equation of State (EoS), one may then describe accurately its
variation throughout the phase diagramme and employ it as an order
parameter. We propose and show that the equilibrated strange-quark
chemical potential of the strongly interacting system is the
sought-for thermodynamic quantity. This suggestion was firstly put
forth earlier in [5].

In sections 2, 3 and 4 we discuss the partition functions for the
strange hadron sector in the HG, chiral QGP and DQM phases,
respectively. Taking into account in an approximate way the
dynamics of the DQM phase, we construct an empirical Equation of
State for this domain. We employ it, together with the known EoS
of the HG and QGP phases, to obtain the strange-quark chemical
potential in a functional form of the temperature and light-quark
chemical potential throughout the phase diagram. With this
relation we study the variation of $\mu_s$, attributing the
changes in the sign and magnitude of the strange-quark chemical
potential to the changes of phase of nuclear matter. In section 5
we summarize several thermal model analyses of experimental
particle yield data from nucleus-nucleus interactions at AGS and
SPS. Among the latter, the S+A interactions at 200 AGeV may be
characterized by the existence of negative strange-quark chemical
potential. In section 6 we predict the values for certain strange
particle ratios and suggest experiments at RHIC at lower energies,
where the baryon number density is finite. Finally, in section 7
we discuss our proposals and come to conclusions whether the S+A
and Pb+Pb interactions at the SPS have entered the deconfined
phase.

\section{Hadron Gas phase}

In the HG phase, the mass spectrum is given in the Boltzmann
approximation by the partition function, $\ln{Z_{HG}}$. We assume
that the hadronic state has attained thermal and chemical
equilibration of three quark flavors (u, d, s):
\begin{eqnarray}
\fl \ln{Z_{HG}}(T,V,\lambda_q,\lambda_s)&=&Z_m+Z_n(\lambda_q^3+
\lambda_q^{-3})
+Z_K(\lambda_q\lambda_s^{-1}+\lambda_q^{-1}\lambda_s)
+Z_Y(\lambda_q^2\lambda_s+\lambda_q^{-2}\lambda_s^{-1})\nonumber\\
&~&~~~~+Z_\Xi(\lambda_q\lambda_s^2+\lambda_q^{-1}\lambda_s^{-2})
+Z_\Omega(\lambda_s^3+\lambda_s^{-3})~~~~~~~~~~(q=u,d)
\end{eqnarray}
The one particle Boltzmann partition function $Z_k$ is given by

\begin{equation*}\label{bolap}
Z_k(V,T)=\frac{VT^3}{2\pi^2}\sum_jg_j\Big(\frac{m_j}{T}\Big)^2K_2\Big
(\frac{m_j}{T}\Big)
\end{equation*}
The fugacity $\lambda_i^k$ controls the quark content of the
k-hadron, i = s, b for strangeness and baryon number [$\lambda_b =
\lambda_q^3 = e^{3\mu_q/T}$], respectively. The summation in Eq.
(2) runs over the resonances of each hadron species with mass
$m_j$, and the degeneracy factor $g_j$ counts the spin and isospin
degrees of freedom of the j-resonance. For the strange hadron
sector, kaons with masses up to 2045 MeV/c2, hyperons up to 2350
MeV/c2 and cascades up to 2025 MeV/c2 are included, as well as the
known $\Omega$-states at 1672 MeV and 2252 MeV. For simplicity, we
assume isospin symmetry in Eq. (1), $\mu_u = \mu_d = \mu_q$.
Strangeness neutrality in strong interactions necessitates:

\begin{equation}
<N_s-N_{\overline{s}}> =
\frac{T}{V}\frac{\partial}{\partial\mu_s}\ln{Z_{HG}(T,V,\lambda_q,\lambda_s)}=0
\end{equation}
which reduces to

\begin{equation}
\fl Z_K(\lambda_q^{-1}\lambda_s-\lambda_q\lambda_s^{-1})
+Z_Y(\lambda_q^2\lambda_s-\lambda_q^{-2}\lambda_s^{-1})+2Z_{\Xi}
(\lambda_q\lambda_s^2-\lambda_q^{-1}\lambda_s^{-2})
+3Z_{\Omega}(\lambda_s^3-\lambda_s^{-3})=0
\end{equation}
This is an important condition as it defines the relation between
light- and strange-quark fugacities in the equilibrated primordial
state. Eq. (4) can be used to derive the true (transverse-flow
independent) temperature of the state, once the fugacities
$\lambda_q$ and $\lambda_s$ are known from experimental strange
particle yield ratios [4].

In the HG phase with finite net baryon number density, the
chemical potentials $\mu_q$ and $\mu_s$ are coupled through the
production of strange hadrons. Due to this coupling, strangeness
conservation does not necessitate $\mu_s = 0$ everywhere in this
phase. In fact, $\mu_s>0$ in the hadronic domain. The condition
$\mu_s=0$ requires $\lambda_s = \lambda_s^{-1} = 1$, and Eq. (4)
becomes [8],

\begin{eqnarray}
\Big[Z_{Y}(\lambda_q+\lambda_q^{-1})-Z_K+2Z_{\Xi}\Big](\lambda_q-
\lambda_q^{-1})=0\nonumber
\end{eqnarray}
The first factor gives the curve $\mu_q$ as a function of T in the
phase diagram, along which $\mu_s = 0$:
\begin{equation}
\mu_q(T)=T \cosh^{-1}\Big(\frac{Z_K}{2Z_Y} -
\frac{Z_{\Xi}}{Z_Y}\Big)
\end{equation}

A more elegant and concise formalism describing the HG phase is
the Strangeness-including Statistical Bootstrap model (SSBM) [9].
It includes the hadronic interactions through the mass spectrum of
all hadron species, in contrast to all other ideal hadron gas
formalisms. The SSBM is valid and applicable only within the
hadronic phase, defining in a determined way the limits of this
phase. The boundary of the hadronic domain is given by the
projection on the 2-dimensional (T,~$\mu_q$) phase diagram of the
intersection of the 3-dimensional bootstrap surface with the
strangeness-neutrality surface ($\mu_s = 0$). Note that the
vanishing of the strange-quark chemical potential on the HG
borderline does not apriori suggest that $\mu_s = 0$ everywhere
beyond this phase. It only states that $\mu_s = 0$ characterizes
the end of the hadronic phase. According to this definition of the
HG boundary, we consider the hadronic phase, described by Eq. (4)
in the ideal HG model, to be terminated at the line where the
strange quark chemical potential vanishes, despite the fact that
Eq. (4) maps the hadronic domain independently of the value of
$\mu_s$. That is, in the ideal HG model we consider the upper
limit of the hadronic phase to be defined by Eq. (5). The relative
position of the two boundaries for the hadron gas domain on the
phase diagram, as defined by the two models, is shown in Fig. 1.
We note that the ideal HG boundary lies about 15 MeV higher than
the SSBM one (for $\mu_q<100 MeV$), due to the absence of
interactions in the former.

\begin{figure}
\begin{center}
\epsfbox{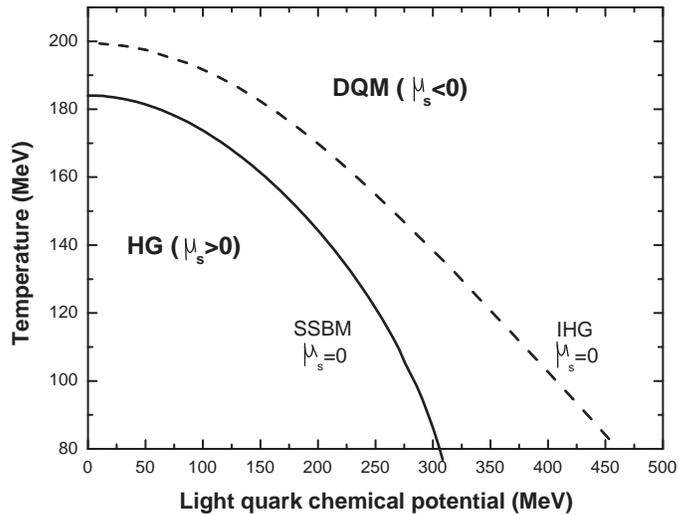}
\end{center}
\caption{The hadronic boundary within the SSBM and IHG models. The
IHG $\mu_s=0$ line lies higher than the SSBM one.}
\end{figure}

\section{Chirally symmetric QGP phase}

In the chirally symmetric QGP region, the partition function for
the current-mass u, d, s-quarks and gluons has the form:

\begin{equation}\label{qgp}
\fl
\ln{Z_{QGP}(T,V,\mu_q,\mu_s)}=\frac{V}{T}\Big[\frac{37}{90}\pi^2T^4+\mu_q^2T^2+
\frac{\mu_q^4}{2\pi^2}+\frac{g_s m_s^{0~2}
T^2}{2\pi^2}(\lambda_s+\lambda_s^{-1})K_2\Big(\frac{m_s^0}{T}\Big)\Big]
\end{equation}
where $m_s^0$ and $g_s$ is the current-mass and degeneracy of the
s-quark, respectively. Strangeness conservation gives $\lambda_s =
\lambda_s^{-1} = 1$, or
\begin{equation}
\mu_s^{QGP}(T,\mu_q)=0
\end{equation}
throughout this phase. Here the two order parameters, the average
thermal Wilson loop $<L>$ and the scalar quark density
$<\overline{\psi}\psi>$, have reached their asymptotic values.
Note that the chirally symmetric quark-gluon plasma phase always
corresponds to a vanishing strange quark chemical potential, even
if we take into account perturbative corrections to the partition
function of Eq. (6). Thus, the chiral symmetry restoration
critical temperature $T_{\chi}$ is not associated with the fact
that we consider, for simplicity, an ideal quark-gluon gas to
describe the QGP.

\section{Deconfined Quark Matter Phase (DQM)}

In formulating our description beyond the hadronic phase, we use
the following picture: The thermally and chemically equilibrated
primordial state at finite baryon number density, produced in
nucleus-nucleus interactions, consists of the deconfined valance
quarks of the participant nucleons, as well as of $q-\overline{q}$
pairs (q=u,d,s), created by quark and gluon interactions. Beyond
but near the HG boundary, $T \geq T_d$, the
correlation-interaction between $q-q$ is near maximum,
$\alpha_s(T) \leq 1$, a prelude to confinement into hadrons upon
hadronization. With increasing temperature, the
correlation-interaction of the deconfined quarks gradually
weakens, $\alpha_s(T) \rightarrow 0$, as colour mobility and
colour charge screening increase. The mass of all (anti)quarks
depend on the temperature and scale according to a prescribed way.
The initially constituent mass decreases with increasing T, and as
the DQM region goes asymptotically into the chirally symmetric QGP
domain, as $T \rightarrow T_{\chi}$, quarks attain current mass.
In this formulation, the equation of state in the DQM region
should lead to the EoS of the hadronic phase, Eq. (1), at $T <
T_d$, and to the EoS of the QGP, Eq. (6), at $T \sim T_{\chi}$.

To construct the empirical EoS in the DQM phase, we use the two
order parameters (for details see Appendix A):
\begin{enumerate}
\item Average thermal Wilson loop, $<L>=e^{-F_q/T}\sim R_d(T)=0
\rightarrow 1$, as $T=T_d \rightarrow T_{\chi}$, describing the
quark deconfinement and subsequent colour mobility, $F_q$ being
the free quark energy. \item Scalar quark density,
$<\overline{\psi}\psi> \sim R_{\chi}(T)= 1 \rightarrow 0$, as $T =
T_d \rightarrow T_{\chi}$, denoting the scaling of the quark mass
with temperature.
\end{enumerate}
Although the Wilson-Polyakov loop is not yet calculated for a
finite density system, we consider that, in the case of non zero
baryochemical potential, it has a smooth temperature dependence,
as suggested in [10\footnote[1]{These finite baryon density
simulations use a canonical partition function and have provided
results in the quenched (heavy quark mass) limit of QCD within the
presence of static quarks.},11], attaining non-limiting values for
a range of temperatures and, therefore, implying the broadening of
the transition region. This permits us to approximate the order
parameter for the finite baryon density case with an arbitrary,
general, yet non-step functional form.

\subsection{Definition of the order parameters}

In order to construct the EoS in the DQM region it is necessary to
employ an analytic functional form of the order parameters. We
have approximated the Wilson loop, obtained from zero density
lattice calculations, with two different functions of temperature
and chemical potential:

\begin{equation}
R_d(T)=\Bigg{\{}
       \begin{array}{l}
    \Theta(T-T_d),~~~~~~~~~~\mu_q=0 \\
    ~~~~\\
    \frac{1}{1+\exp[-\alpha(T-\beta)]},~~~~\mu_q>0
  \end{array}
  \end{equation}
and
\begin{equation}
R_d(T)=\Bigg{\{}
       \begin{array}{l}
    \Theta(T-T_d),~~~~~~~~~\mu_q=0 \\
    ~~~~\\
    c\Big(\frac{T-T_d}{T}\Big)^{\nu},~~~~~~~~\mu_q>0
  \end{array}
  \end{equation}
In Eq. (7) a Fermi-type function is used for finite $\mu_q$, where
the parameters $\{\alpha,~\beta\}$ control the difference of the
critical temperatures $T_d$ and $T_{\chi}$, i.e., when $R_d \sim
0$ and $R_d \sim 1$, respectively. They can be chosen arbitrarily
in order to obtain realistic compatibility with the lattice QCD
results. In Eq. (8) the order parameter is described using a more
common, in the theory of critical phenomena, function near the
critical point $T_d$, where the critical exponent $\nu$ can be
chosen according to universality class arguments or arbitrarily.
The parameter $c$ simply controls the values of $R_d(T)$, so that
asymptotically $R_d(T) \rightarrow 1$ as $T \rightarrow T_{\chi}$.
Note that the exact expression for the considered order parameter
$R_d(T)$ is not of crucial importance for this qualitative
description (see Appendix A). Figures 2, 3 show the approximated
order parameter for different values of $\{\alpha,~\beta\}$ and of
the critical exponent $\nu$, respectively, for the finite chemical
potential case, $\mu_q > 0$. In Figure 4 we show the two
approximated order parameters, where for simplicity we have taken
$R_{\chi}(T) \equiv 1 - R_d(T)$ and used Eq. (8) with the
parameterization $\{\alpha=0.03,~\beta=268\}$.

\begin{figure}[ht]
\begin{center}
\epsfbox{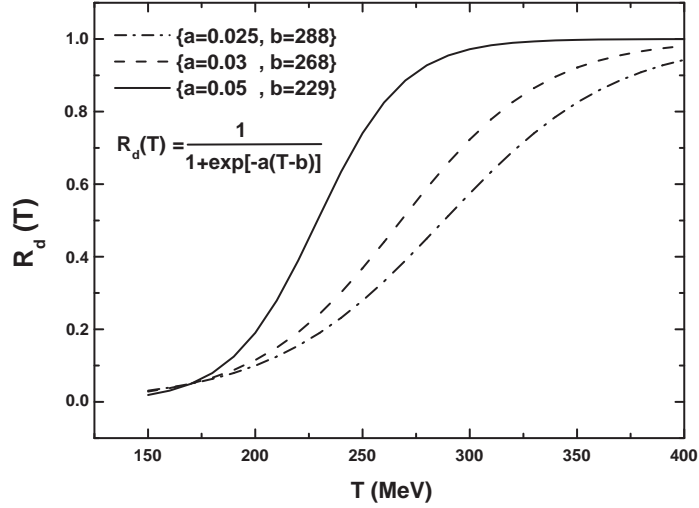}
\end{center}
\caption{Approximated finite density Polyakov loop $R_d(T)$,
obtained from Eq.(8) for different values of the model parameters
$\{\alpha, \beta\}$.}
\end{figure}

\begin{figure}[ht]
\begin{center}
\epsfbox{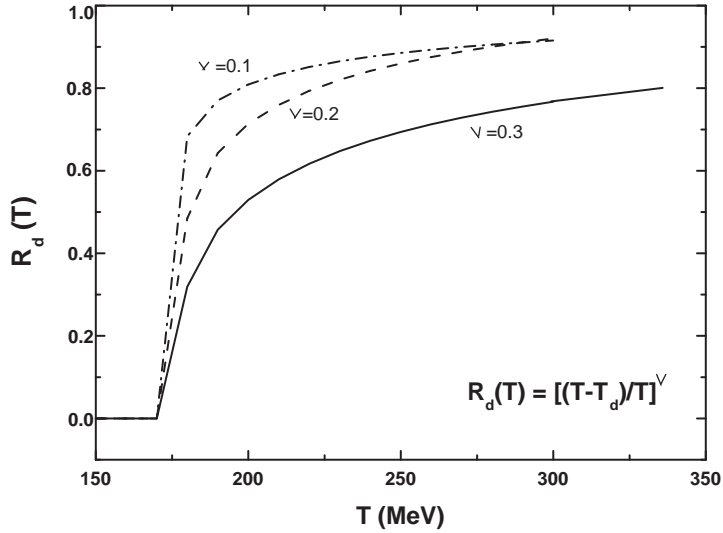}
\end{center}
\caption{Approximated finite density Polyakov loop $R_d(T)$,
obtained from Eq.(9) for different values of the model parameter
$\nu$.}
\end{figure}

\begin{figure}[tb]
\begin{center}
\epsfbox{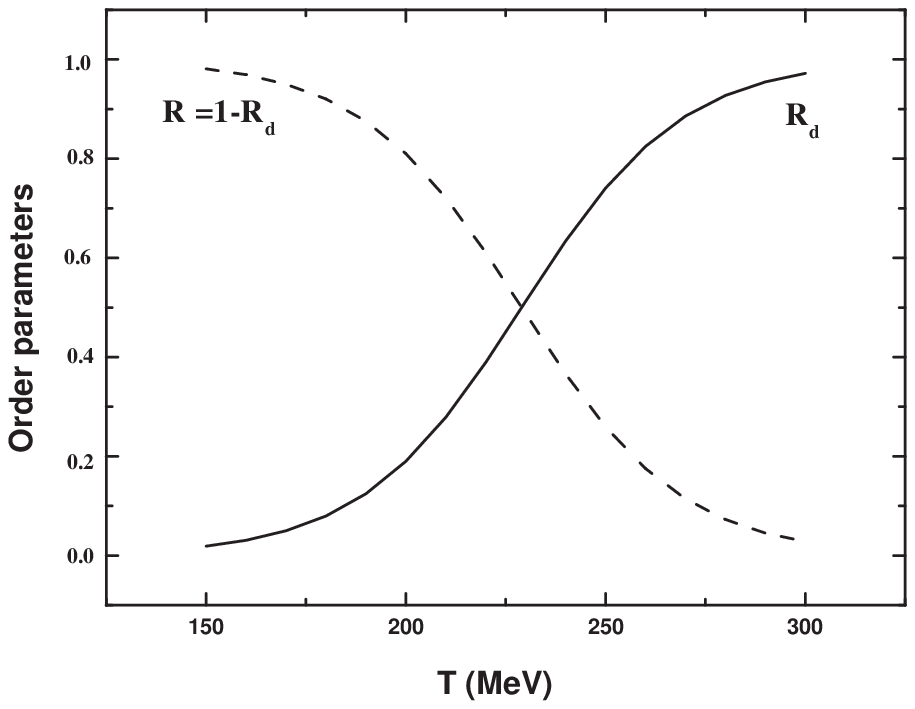}
\end{center}
\caption{Plot of the approximated order parameters $R_d(T)$ and
$R_{\chi}(T)=1-R_d(T)$ using Eq.(7) with $\alpha=0.03$ and
$\beta=268$.}
\end{figure}

\subsection{The DQM Equation of State}

We assume that above $T_d$ the deconfined quarks retain a degree
of correlation, resembling "hadron-like" states, since $1 >
\alpha_s(T)>0$. The diminishing of this correlation-interaction,
as a result of progressive increase of colour mobility, is
approximated by the factor: $[1-R_d(T)] = 1 \rightarrow 0$, as $T
= T_d \rightarrow T_{\chi}$. Note that effectively $[1-R_d(T)]
\sim \alpha_s(T)$ in the DQM region, the temperature dependence of
the running coupling constant [12]. To account for the "effective
mass" of the state as a function of temperature, we assume the
mass of the quarks to decrease from the constituent value and
reach the current-mass as $T \rightarrow T_{\chi}$. The quark mass
scales with temperature as:

\begin{equation}
m_q^*(T)=R_{\chi}(T)(m_q-m_q^0)+m_q^0
\end{equation}
where $m_q$ and $m_q^0$ are the constituent and current quark
masses, respectively, ($m_u^0 \sim 5~MeV,~m_d^0 \sim 9~MeV,~m_s^0
\sim 170~MeV$). Similarly, the effective hadron-like mass scales
as:
\begin{equation}
m_i^*(T)=R_{\chi}(T)(m_i-m_i^0)+m_i^0
\end{equation}
where $m_i$ is the hadron mass in the hadronic phase and $m_i^0$
is equal to the sum of the hadron's quarks current-mass ($m_K^0
\sim 175 MeV,~m_Y^0 \sim 185 MeV,~m_X^0 \sim 350 MeV$ and
$m_{\Omega}^0 \sim 510 MeV$). In the EOS, the former scaling is
employed in the mass-scaled partition function $\ln{Z_{QGP}^*}$,
whereas the latter in the mass-scaled partition function
$\ln{Z_{HG}^*}$, which accounts for the produced hadron species.
Note that this mass-scaling is effectively equivalent to the one
given in the
Nambu - Jona-Lasinio (NJL) formalism [13] (see Appendix B).\\
Employing the described dynamics, we construct an empirical
partition function of the DQM phase:
\begin{equation}
\fl
\ln{Z_{DQM}(V,T,\lambda_q,\lambda_s)}=[1-R_d(T)]\ln{Z_{HG}^*(V,T,
\lambda_q,\lambda_s)}+R_d(T)\ln{Z_{QGP}^*(V,T,\lambda_q,\lambda_s)}
\end{equation}
The proposed intermediate DQM phase should not be confused with a
mixed hadronic-partonic phase. The partition function of Eq. (11)
is written as a linear combination of the HG and QGP mass-scaled
partition functions. It is a realistic and convenient
approximation, which satisfies the general property of the
partition function, describing asymptotically confinement and
chiral symmetry restoration. In the intermediate temperature
regime of the DQM phase, neither pure hadronic bound states, nor
chirally symmetric QGP clusters can be found. In Eq. (11) the
factor $[1-R_d(T)]$ describes the weakening of the
correlation-interaction of the deconfined quarks constituting the
"hadron-like" entities and   gives the mass-scaling of these
entities with increasing temperature. In the second term, the
factor $R_d(T)$ defines the rate of colour mobility, whereas
$\ln{Z_{QGP}^*}$ represents the state as it approaches the QGP
region. Thus, at $T = T_d$, the EoS of the DQM region goes over to
the corresponding in the HG phase and at $T \sim T_{\chi}$, to the
EoS in the chirally symmetric QGP region.\\Strangeness neutrality
in the DQM phase leads to the EoS:
\begin{eqnarray}
\fl
[1-R_d(T)]\Big[Z_K^*(\lambda_s\lambda_q^{-1}-\lambda_q\lambda_s^{-1})+
Z_Y^*(\lambda_s\lambda_q^2-\lambda_s^{-1}\lambda_q^{-2})
+2Z_{\Xi}^*(\lambda_s^2\lambda_q-
\lambda_s^{-2}\lambda_q^{-1})\nonumber\\
+3Z_{\Omega}^*(\lambda_s^3-\lambda_s^{-3})\Big]+R_d(T)g_s
m_s^{*2}K_2\Big(\frac{m_s^*}{T}\Big)(\lambda_s-\lambda_s^{-1})=0
\end{eqnarray}
For given $\mu_q$, Eq. (12) defines the variation of the
strange-quark chemical potential with temperature in the DQM
domain. Combining Eq's (4, 12) we obtain the change of the
strange-quark chemical potential with temperature in the entire
phase diagram.

\subsection{Zero chemical potential, $\mu_q=0$}

In the case of vanishing baryon chemical potential, quarks attain
current mass instantaneously upon deconfinement, due to the fact
that $R_{\chi}(T)$ becomes a step function with
$R_{\chi}(T,\mu_q=0) = 0$, for temperatures $T > T_d$, Eq's (7),
(8). Consequently, the chirally symmetric QGP phase follows
immediately after the HG phase, as can be seen from Eq. (12) and
the partition function beyond the HG region is given by Eq. (6).
The deconfinement and chiral symmetry transitions coincide, Td =
T÷, and the DQM phase with negative strange-quark chemical
potential vanishes. Due to vanishing baryon chemical potential,
the EoS (12) leads to $\mu_s = 0$ for all temperatures. In this
case, the strange-quark chemical potential cannot provide a signal
for any phase transition. Figure 5 exhibits this effect, showing
the variation of $\mu_s$ with temperature for different values of
the light-quark chemical potential $\mu_q$.

\begin{figure}[tb]
\begin{center}
\epsfbox{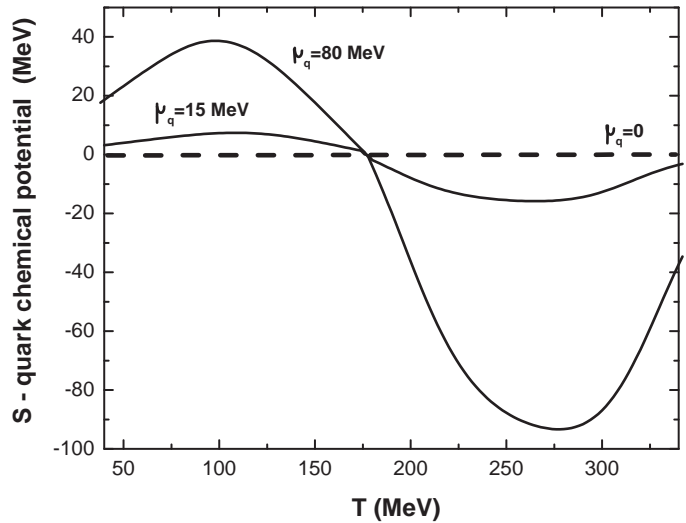}
\end{center}
\caption{Variation of $\mu_s$ with T for various values of
$\mu_q$. For zero $\mu_q$, $\mu_s$ vanishes for all temperatures.}
\end{figure}

\subsection{Finite chemical potential, $\mu_q > 0$}

The situation for a non-vanishing baryon chemical potential is not
yet known in terms of lattice calculations. Within this model we
can easily extend our analysis for a system of finite and fixed
$\mu_q$ according to Eq. (12). In Fig. 6 the strange-quark
chemical potential maps the QCD phase diagram as a function of
$T$, for $\mu_q/T= 0.45$. We observe that $\mu_s$ attains positive
values in the HG phase, as a result of the coupling between
$\mu_q$ and $\mu_s$ in strange hadrons. It approaches zero as the
hadron density reaches its asymptotic Hagedorn limit at the end of
the hadronic phase, where a phase transition to partonic matter
takes place, at the deconfinement temperature $T_d$ [14]. At this
temperature, $\mu_s(T_d) = 0$, signifying the vanishing of this
coupling in hadrons. In the DQM region it grows strongly negative,
where an effective correlation - progressively weakening - remains
in effect among the deconfined, but interacting quarks with
non-current mass. Finally $\mu_s$ returns to zero as the QGP phase
is approached, with restored chiral symmetry (current-mass quarks)
and $\alpha_s$ approaching zero.\\Figure 7 exhibits the variation
of $\mu_s/T$ as a function of $T$ throughout the 3-region phase
diagram for $\mu_q = 0.45T$. We have assumed that the
deconfinement temperature is $T_d \sim 176~MeV$ at $\mu_q \sim
80~MeV$, as given by the SSBM hadronic phase boundary [15]. We
observe that:

\begin{figure}[tb]
\begin{center}
\epsfbox{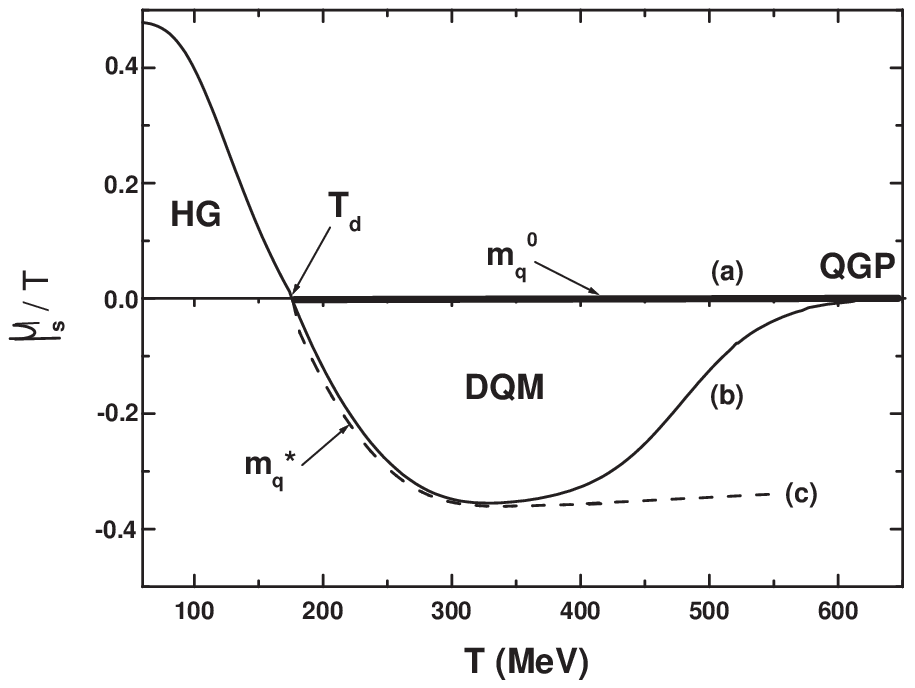}\\\footnotesize{\textbf{Figure 7.} Variation
of $\mu_s/T$ with temperature in the DQM region for different
considerations of the parameters of the EoS, eq. (12).}
\end{center}
\end{figure}

a) If the deconfined-quark mass were to attain its current value
($m_q=m_q^0$) beyond $T_d$, then $\mu_s$ would be zero for
$T>T_d$, curve (a).

b) If $0 < R_d < 1$ and $m_q^0 < m_q^*(T) < m_q$, then the EoS
(12), which includes mass-scaled "hadron-like" states, as well as
(anti)quarks with scaled mass, gives large negative values for
$\mu_s/T$ in the DQM phase. After reaching a minimum, $\mu_s/T$
returns asymptotically to zero as the QGP phase is approached at
$T_{\chi} \sim (2 - 4) T_d$, depending in general on $\mu_q$, Fig.
7, curve (b).

c) Scaling alone of the effective hadron masses produces large
negative strange-quark chemical potential beyond the HG phase,
saturating at high temperature without ever returning to zero at
the QGP phase, Fig. 7, curve (c).

The fact that we obtain a chiral symmetry restoration temperature
$T_{\chi} \sim (2 - 4) T_d$ for $\mu_q \sim 80~MeV$, is a result
of the specific parameterization used for the order parameter. The
DQM phase exists for $0 < R_d < 1$, so it is actually the
condition $R_d(T_{\chi}) = 1$ that determines the end of DQM at $T
\rightarrow T_{\chi}$. In general, $T_{\chi} = k(\mu_q)T_d$, where
$k(\mu_q>0)
> 1$ and $k(\mu_q=0) = 1$. However, finite baryon chemical potential
will be always associated with the existence of negative $\mu_s$
and therefore $T_{\chi} > T_d$, meaning that the two phase
transitions, deconfinement and chiral, are apart. This is an
interesting result, which is connected with the existence of
negative strange-quark chemical potential. Note that there is no
known argument from QCD that the two transitions occur at the same
temperature for finite baryon density.

On the basis of the above observations, we propose that the change
in the sign of the strange-quark chemical potential, from positive
in the HG phase to negative in the DQM, defines uniquely and
precisely the phase transition to quark-deconfinement at finite
baryon density. The experimental observation of negative $\mu_s$
would be an evidence for the existence of the DQM region and
consequently of the fact that chiral symmetry restoration follows
deconfinement ($T_d < T_{\chi}$). We note that the sign of
$\mu_s$, positive in the HG, negative in the DQM and zero in the
QGP domains, is independent of the particular form of the
parameters $R_d$ and $R_{\chi}$ used in the EoS and unique in each
region (see Appendix A). It can be considered and used as a
potentially experimentally accessible order parameter of the
deconfinement phase transition. This signature is independent of
assumptions and ambiguities regarding interaction mechanisms and
matter media effects, as well as weakness and uncertainties in the
predictive power of models, as is the case with other proposed
signs for deconfinement: $J/\psi$ suppression, resonance shift and
broadening, strangeness enhancement, etc. In the present
calculation, however, the magnitude and shape of the negative
chemical potential curve as a function of temperature should be
taken in a qualitative manner, due to the empirical treatment of
the dynamics in the DQM phase. A detailed, quantitative treatment
of the DQM EoS will require the use of a three-flavor effective
Lagrangian in the NJL formalism [16].

\section{Experimental Data}

Data from nucleus-nucleus interactions, obtained by experiments
E802 [17] at AGS and NA35 [18], NA49 [19] at SPS have been
analyzed in terms of several statistical-thermal models: the
Strangeness-including Statistical Bootstrap Model, SSBM [9, 15,
20-22] and others employing the canonical and grand-canonical
formalisms [23-27]. The data analysis is made, of course, within
the inevitable limitations of all thermal-statistical models and
the assumption of a thermalized and equilibrated matter. There are
proposals for possible non-equilibrium effects, however thermal
models seem to be in surprisingly well consistency with the
experimental data. Table 1 summarizes the results of these
analyses, from which the quantities $T$, $\mu_q$ and $\mu_s$ have
been deduced. Observe that all interactions, studied so far,
exhibit positive $\mu_s$ and therefore there does not yet exist a
confident experimental confirmation of negative values for the
strange-quark chemical potential. Note that in [23,25,26] the
negative $\mu_S$ refers to the strangeness chemical potential,
which is related to the strange-quark chemical potential by:
$\mu_S = \mu_s+\mu_q$. This fact led to a misinterpretation in
[28]. However the sulfur-induced interactions may deserve a more
cautious examination, as suggested by the SSBM analysis. In
contrast to other thermal-statistical models, the Statistical
Bootstrap Model (SBM) of Hagedorn [29] incorporates the hadronic
interactions in the EoS through the bootstrap equation. The
development of the SBM with the inclusion of Strangeness (SSBM)
has been employed in the analysis of nucleus-nucleus interactions
at the SPS. The SSBM analysis for the S+S interaction [15] has
shown that this equilibrated state is situated mostly (75\%)
outside the hadronic phase, whereas the S+Ag [21] is just on the
deconfinement line. In addition, the calculations have pointed out
a large ($\sim$30\%) entropy enhancement of the experimental data
compared to the model, an effect observed also by other
calculations [23-25]. This enhancement may be attributed to
contributions from the DQM phase with many liberated new partonic
degrees of freedom.
\begin{figure}[htb]
\begin{center}
\epsfbox{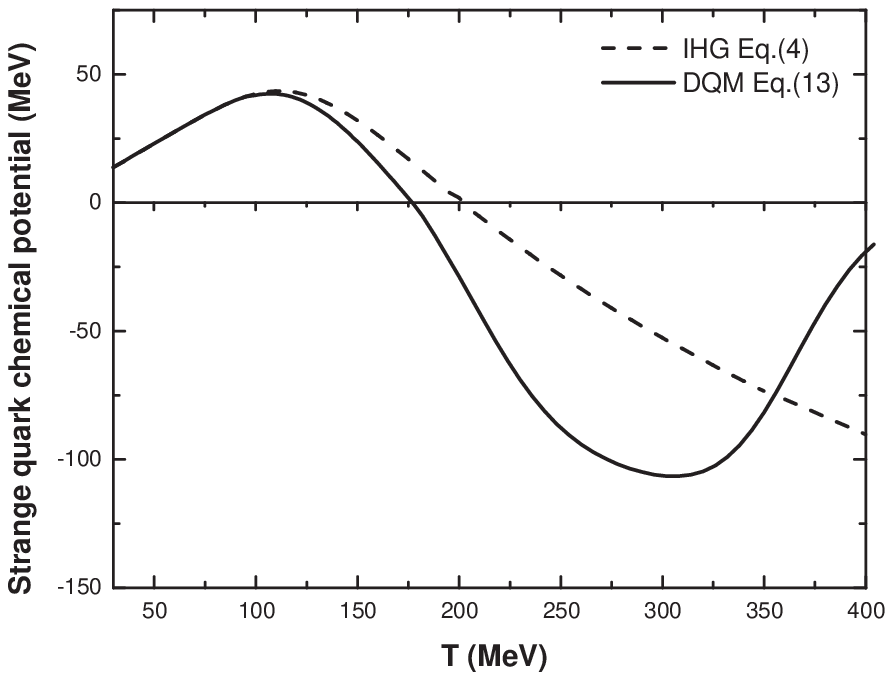}\\\footnotesize{\textbf{Figure 8.}Variation of
the strange quark chemical potential with the temperature within
the IHG formalism and the proposed DQM. The change in the sign of
$\mu_s$ is realized at a higher temperature for the IHG model.}
\end{center}
\end{figure}

\begin{figure}[htb]
\begin{center}
\epsfbox{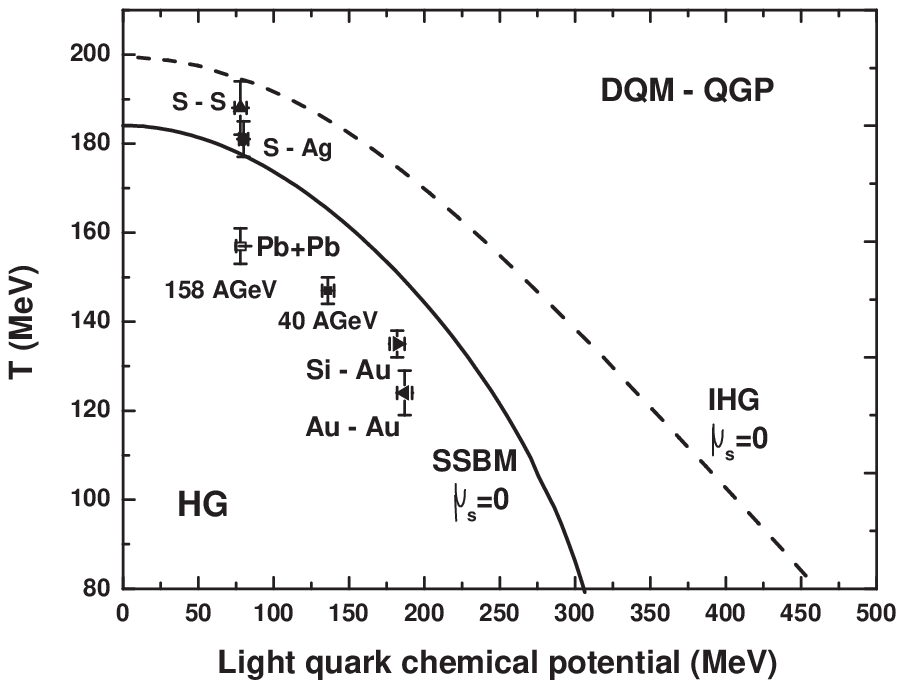}\\\footnotesize{\textbf{Figure 9.}Phase diagram
with the SSBM deconfinement line and the location (T, $\mu_q$
average values) of several interactions, deduced from the analysis
with thermal models.}
\end{center}
\end{figure}

\begin{table}[t]
  \centering
\small{
\begin{tabular}{lllll}

\multicolumn{5}{l}{\textbf{Table 1}. Deduced values for T,
$\mu_q$, $\mu_s$ from several thermal} \\
\multicolumn{5}{l}{models and fits to experimental data for
several interactions.}\\ \hline \hline
\multicolumn{5}{c}{\textbf{Interaction/Experiment}}\\ \hline
\multicolumn{5}{c}{Si+Au(14.6 AGeV)/E802}\\
   & Reference[28] & Reference[26] & Mean & \\
  T(MeV) & 134$\pm$6 & 135$\pm$4 & 135$\pm$3 & \\
  $\mu_q$(MeV) & 176$\pm$12 & 194$\pm$11 & 182$\pm$5 &  \\
  $\mu_s$(MeV) & 66$\pm$10 & & 66$\pm$10 & \\
  \multicolumn{5}{c}{Au+Au(11.6 AGeV)/E802}\\
   & Reference[28] & Reference[26] & Mean & \\
  T(MeV) & 144$\pm$12 & 121$\pm$54 & 124$\pm$5 & \\
  $\mu_q$(MeV) & 193$\pm$17 & 186$\pm$5 & 187$\pm$5 &  \\
  $\mu_s$(MeV) & 51$\pm$14 & & 51$\pm$14 & \\
  \multicolumn{5}{c}{Pb+Pb(158 AGeV)/NA49}\\
  & Reference[28] & Reference[26] & Reference[22] & Mean \\
  T(MeV) & 146$\pm$9 & 158$\pm$3 & 157$\pm$4 & 157$\pm$3\\
  $\mu_q$(MeV) & 74$\pm$6 & 79$\pm$4 & 81$\pm$7 & 78$\pm$3 \\
  $\mu_s$(MeV) & 22$\pm$3 & & 25$\pm$4 & 23$\pm$2\\
  \multicolumn{5}{c}{Pb+Pb(40 AGeV)/NA49$^*$}\\
  & Reference[28] & Reference[32] & Mean & \\
  T(MeV) & 147$\pm$3& 150$\pm$8 & 149$\pm$9 & \\
  $\mu_q$(MeV) & 136$\pm$4 & 132$\pm$7 & 134$\pm$8 & \\
  $\mu_s$(MeV) & 35$\pm$4 & & & \\
  \multicolumn{5}{c}{S+S(200 AGeV)/NA35}\\
  & Reference[23] & Reference[25] & Reference[24] & Mean \\
  T(MeV) & 182$\pm$9 & 181$\pm$11 & 202$\pm$13 & 188$\pm$6\\
  $\mu_q$(MeV) & 75$\pm$6 & 73$\pm$7 & 87$\pm$7 & 78$\pm$4 \\
  $\mu_s$(MeV) & 14$\pm$4 & 17$\pm$6 & & 16$\pm$7\\
  \multicolumn{5}{c}{S+Ag(200 AGeV)/NA35}\\
   & Reference[23] & Reference[25] & Reference[24] & Mean \\
  T(MeV) & 180$\pm$3 & 179$\pm$8 & 185$\pm$8 & 181$\pm$4\\
  $\mu_q$(MeV) & 79$\pm$4 & 81$\pm$6 & 81$\pm$7 & 80$\pm$3 \\
  $\mu_s$(MeV) & 14$\pm$4 & 16$\pm$5 & & 16$\pm$8\\
   & & & & \\
  \multicolumn{5}{l}{*NA49 private communication}\\
  \hline \hline
  \end{tabular}
}
 \end{table}

The fact that beyond the HG phase the deconfined quarks retain a
degree of correlation, resembling "hadron-like" states, allows the
thermal models [23-25] to describe the state adequately using a
hadron gas EoS. These models do not sense the HG phase limit and
treat the deconfined state as consisting of colorless hadrons.
However, an ideal Hadron Gas model with no interactions exhibits
strong deviations from the SSBM as we approach the critical region
($T \sim 175~MeV$), where the sulfur-induced interactions are
approximately situated. Within the IHG model the condition $\mu_s
= 0$ is satisfied for a higher temperature $T_0 \sim 200~MeV$,
extending the HG phase to a larger region in the phase diagram
(see Fig. 1). Consequently, the negative strange-quark chemical
potential, a characteristic of the region beyond the HG phase,
appears at higher temperatures also for the ideal HG model, Fig.
8. This is the reason why $\mu_s > 0$ in the analysis of [23, 25],
although a temperature above the deconfinement (according to the
SSBM) is obtained. If the IHG curve is shifted in a way that would
approximate the SSBM boundary, then negative $\mu_s$ should be
obtained for the S+A interactions. Note that this effect is not
important in the study of the other A+A interactions, because in
their case the equilibrated states are situated far below the
critical temperature and therefore the HG model is very close to
the analysis by the SSBM. This can be seen in Table 1, where the
analysis of the experimental data within the two models is almost
identical for extracted temperatures below 160 MeV. For the $Pb +
Pb$ interaction, the SSBM analysis [22] has shown the system to be
located well within the hadronic phase. In addition, thermal model
calculations [22, 26] find no entropy enhancement for this
interaction. Figure 9 shows the 3-flavor phase diagram with the
SSBM maximally extended\footnote[2]{The maximally extended HG
phase limit is defined for $T_0(\mu_q=0) \sim 183~MeV$, which is
the maximum temperature for non-negative $\mu_s$ in the HG domain.
Recent lattice QCD calculations give $T_0 \sim 175~MeV$ for [2+1]
quark flavors [7].} deconfinement line, the IHG $\mu_s = 0$ line
and the location of the state ($T,~\mu_q$ mean values) for several
interactions, Table 1. Observe that the sulfur-induced
interactions at 200 $AGeV$, are situated beyond the SSBM
deconfinement line, whereas all others are well within the
hadronic phase.

\section{Predictions for particle yield ratios at RHIC}

New data obtained at RHIC with the $Au+Au$ interaction at
$\sqrt{s} = 130~AGeV$ (see a compilation of data in [30]) show
that at midrapidity the light quark fugacity is $\lambda_q \sim
1.08$, indicating very small quark chemical potential. This has an
effect on the strange-quark chemical potential, making it close to
zero throughout the 3-region phase diagram (see Fig. 5). In such
case of minimal net baryon density, our model cannot be applied,
as discussed in Section 4.3. We have, therefore, calculated the
particle yield ratios $\Omega^+/\Omega^-$ and $\Xi^+/\Xi^-$ for
finite baryon density, obtained in $Au+Au$ interactions at
$\sqrt{s} = 20 - 90~AGeV$. The quantities needed for these
calculations, $\mu_q/T$ and $T$ of the equilibrated state, were
obtained by fitting all available corresponding values, obtained
for equilibrated states in nucleus-nucleus collisions from SIS to
SPS [26, 27] and extrapolating to RHIC energies. The data were
fitted as a function of ($\sqrt{s}/participant$).
\begin{figure}[htb]
\begin{center}
\epsfbox{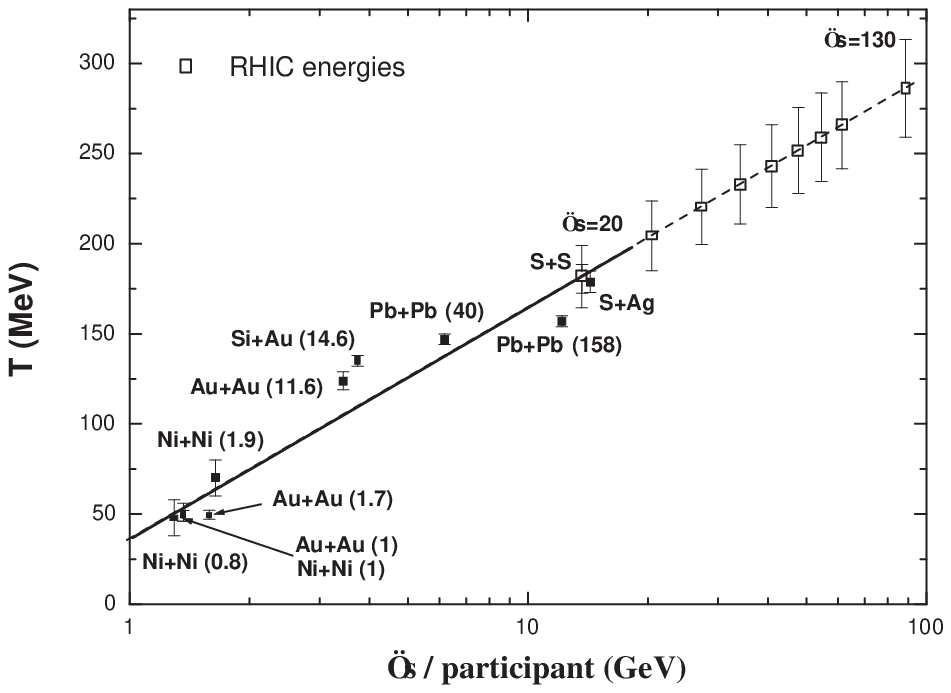}\\\footnotesize{\textbf{Figure 10.}Fitted
temperatures as a function of ($\sqrt{s}/participant$) and
extrapolation to RHIC energies $20<\sqrt{s}<130 AGeV$. The energy
in parenthesis is in units of AGeV.}
\end{center}
\end{figure}
\begin{figure}[htb]
\begin{center}
\epsfbox{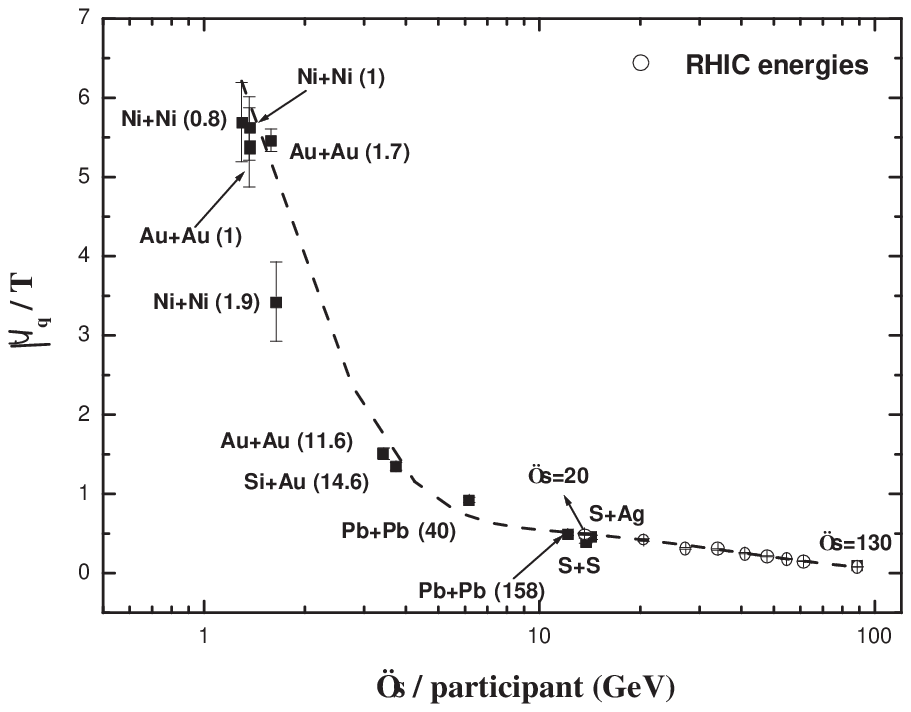}\\\footnotesize{\textbf{Figure 11.}Fit to the
$\mu_q/T$ values of several interactions and extrapolation to RHIC
energies $20<\sqrt{s}<130 AGeV$. The energy in parenthesis is in
units of AGeV.}
\end{center}
\end{figure}
\begin{figure}[htb]
\begin{center}
\epsfxsize=300pt \epsfysize=240pt
\epsfbox{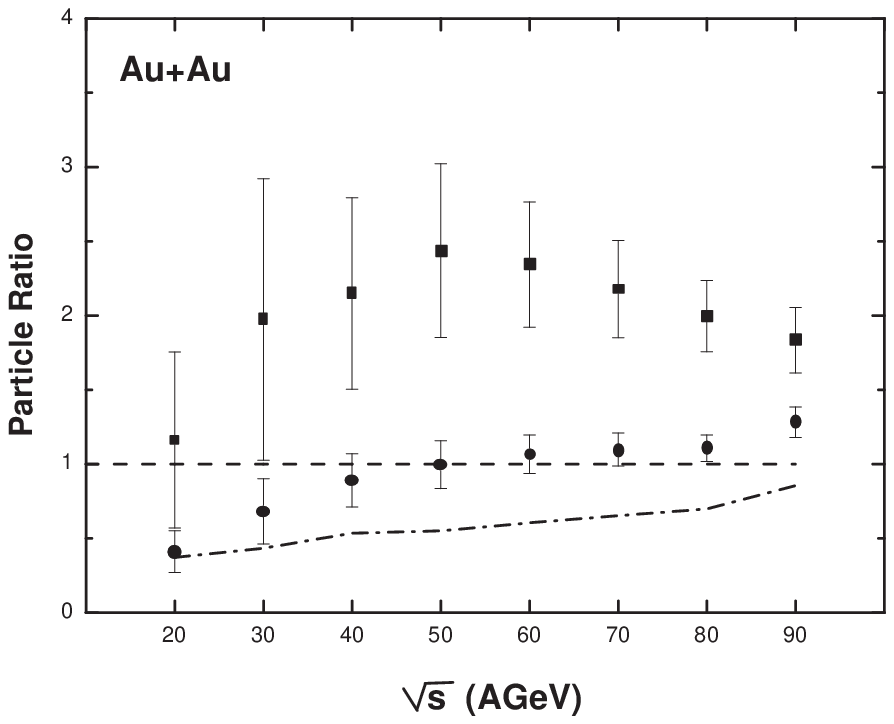}\\\footnotesize{\textbf{Figure
12.}Predicted particle yield ratios $\Omega^+/\Omega^-$
[$\blacksquare$] and $\Xi^+/\Xi^-$ [$\bullet$] for a finite baryon
density region, obtained in $Au+Au$ interactions at $\sqrt{s}$ =
20 - 90 AGeV. The lines correspond to the case $\mu_s=0$ for the
ratio $\Omega^+/\Omega^-$ [--~--~--] and $\Xi^+/\Xi^-$ [-- $\cdot$
-- $\cdot$].}
\end{center}
\end{figure}
\begin{table}[t]
  \centering
\small{
\begin{tabular}{cccccc}
\multicolumn{6}{l}{\textbf{Table 2}. The extrapolated values for
the temperature T and the ratio $\mu_q/T$ obtained}\\
\multicolumn{6}{l}{for the Au+Au interaction at RHIC, as well as
predictions of the empirical EoS for $\mu_s/T$} \\
\multicolumn{6}{l}{and two strange-particle ratios. Table (a)
exhibits the results for $\{\alpha=0.03,~\beta=268\}$ and}\\
\multicolumn{6}{l}{(b) for $\{\alpha=0.05,~\beta=229\}$ in Eq.
(7).}\\ & & & & & \\ (a) & & & & & \\ \hline $\sqrt{s}(AGeV)$ &
T(MeV) & $\mu_q/T$ & $\mu_s/T$ & $\Omega^+/\Omega^-$ &
$\Xi^+/\Xi^-$\\ \hline 20 & 181.76 $\pm$ 17.30 & 0.49 $\pm$ 0.02 &
-0.03 $\pm$ 0.17 & 1.20 $\pm$ 1.22 & 0.42 $\pm$ 0.28 \\ 30 &
204.40 $\pm$ 19.35 & 0.42 $\pm$ 0.01 & -0.20 $\pm$ 0.10 & 3.24
$\pm$ 1.94 & 0.95 $\pm$ 0.38
\\ 40 & 220.45 $\pm$ 20.85 & 0.31 $\pm$ 0.01 & -0.21 $\pm$ 0.05 &
3.50 $\pm$ 1.05 & 1.23 $\pm$ 0.25 \\ 50 & 232.91 $\pm$ 22.02 &
0.30 $\pm$ 0.01 & -0.22 $\pm$ 0.05 & 3.77 $\pm$ 1.13 & 1.33 $\pm$
0.27
\\ 60 & 243.10 $\pm$ 23.06 & 0.25 $\pm$ 0.01 & -0.19 $\pm$ 0.03 &
3.18 $\pm$ 0.57 & 1.30 $\pm$ 0.16 \\ 70 & 251.70 $\pm$ 23.81 &
0.21 $\pm$ 0.01 & -0.16 $\pm$ 0.01 & 2.68 $\pm$ 0.16 & 1.26 $\pm$
0.05
\\ 80 & 259.15 $\pm$ 24.53 & 0.18 $\pm$ 0.01 & -0.14 $\pm$ 0.02 &
2.29 $\pm$ 0.19 & 1.21 $\pm$ 0.07 \\ 90 & 265.72 $\pm$ 24.16 &
0.15 $\pm$ 0.01 & -0.11 $\pm$ 0.02 & 1.99 $\pm$ 0.24 & 1.06 $\pm$
0.04\\ 130 & 286.25 $\pm$ 27.16 & 0.07 $\pm$ 0.01 & -0.05 $\pm$
0.02 & 1.37 $\pm$ 0.08 & 1.21 $\pm$ 0.07 \\ \hline \vspace{5pt}\\
(b) & & & & &\\ \hline $\sqrt{s}(AGeV)$ & T (MeV)$^*$ &
$\mu_q/T~^*$ & $\mu_s/T$ & $\Omega^+/\Omega^-$ & $\Xi^+/\Xi^-$
\\ \hline 20 & & & -0.03 $\pm$ 0.09 & 1.16 $\pm$ 0.60 & 0.41 $\pm$
0.14\\ 30 & & & -0.11 $\pm$ 0.08 & 1.97 $\pm$ 0.95 & 0.68 $\pm$
0.22\\ 40 & & & -0.13 $\pm$ 0.05 & 2.15 $\pm$ 0.64 & 0.89 $\pm$
0.18\\ 50 & & & -0.15 $\pm$ 0.04 & 2.43 $\pm$ 0.58 & 1.00 $\pm$
0.16\\ 60 & & & -0.14 $\pm$ 0.03 & 2.34 $\pm$ 0.42 & 1.07 $\pm$
0.13\\70  & & & -0.13 $\pm$ 0.03 & 2.18 $\pm$ 0.33 & 1.10 $\pm$
0.11\\ 80 & & & -0.12 $\pm$ 0.02 & 2.00 $\pm$ 0.24 & 1.11 $\pm$
0.09\\ 90 & & & -0.10 $\pm$ 0.02 & 1.83 $\pm$ 0.22 & 1.28 $\pm$
0.10\\ 130 & & & -0.05 $\pm$ 0.01 & 1.38 $\pm$ 0.05 & 0.92 $\pm$
0.03\\ \hline \multicolumn{6}{l}{*The same fitted T, $\mu_q/T$
values as in (a).}
\end{tabular}
}
 \end{table}
Figures 10, 11 show the fitted temperatures and $\mu_q/T$ and
their extrapolation to higher energies, respectively. Table 2
contains the extrapolated values for $T$ and $\mu_q/T$,
corresponding to equilibrated primordial states (not chemical
freeze out), as well as the order of magnitude predictions of our
empirical EoS in DQM for $\mu_s/T$ and for the two particle
ratios, for two different $R_d$ curves. A worth mentioning result
of the extrapolation is that, at $\sqrt{s} = 130~AGeV$, a
primordial fireball temperature $T \sim O(280~MeV)$ is found,
which is much higher than the freeze out temperature $T \sim
175~MeV$, obtained in [30]. It may be that in the region of
minimal baryon chemical potential, the hadronization process of a
deconfined state is continuous and smooth and thus the primordial
thermodynamic quantities are not conserved. It should also be kept
in mind that the available RHIC data span a very small rapidity
range ($\Delta y < \pm 1$), making unreliable the extraction of
values for global thermodynamic quantities from these fits. Figure
12 shows the range of the predicted particle ratios and the
maximum values of the ratios in the case where $\mu_s$ does not
become negative, but stays zero after deconfinement. Note that
negative $\mu_s$ means that the particle yield ratio
$\Omega^+/\Omega^-=e^{-6\mu_s/T}$ becomes larger than unity. We
find, within the variance of the extrapolation and the different
$R_d$ curves that, at about $\sqrt{s} \sim 50~AGeV$, both ratios
attain their largest values $\sim$ (2 - 4) and $\sim$ (1 - 1.5)
for $\Omega^+/\Omega^-$ and $\Xi^+/\Xi^-$, respectively, compared
to the maximum value of 1 and 0.55, respectively, for the case
where $\mu_s=0$, Table 2. The observation of the above predictions
is contingent on the assumption that the $T$, $\mu_q$, $\mu_s$
values of the primordial equilibrated state are conserved until
hadronization and thus can be deduced from the particle yield
ratios (see discussion below).

\section{Summary and Discussion}

We have shown that the existence of an intermediate region of
deconfined, massive and correlated quarks, in-between the hadronic
and chiral quark-gluon phases, is realistic and within the results
of recent lattice and effective Lagrangian calculations. We have
constructed an empirical EoS for the DQM phase, in terms of the
order parameters and the mass-scaled partition functions of the HG
and QGP phases, from which a relation for the strange-quark
chemical potential is obtained in terms of $\mu_q$ and $T$. This
empirical EoS includes mass-scaled strange hadrons, approximating
the effects of the progressive decrease of the
interaction-correlation with increasing temperature, as well as a
mass-scaled QGP term. In effect, it describes realistically the
gross features of the dynamics and characteristics of the DQM
phase. The EoS indicates that an unambiguous and concise
characteristic observable of this phase is the large negative
strange-quark chemical potential. It also shows that $\mu_s$ may
possibly be considered and used as an experimentally accessible
order parameter of the deconfinement phase transition, the only
such parameter proposed up to now.

Negative chemical potential appears also in condensed matter
systems. For example, in a transition between weekly coupled
Cooper pairs, with $\mu > 0$ and the usual BCS superconducting gap
$|\Delta k|$, and the strongly coupled diatomic pairs, with $ \mu
< 0$ and the corresponding gap $(\Delta k^2 + \mu^2)^{1/2}$,
representing an insulating system [31]. The analogy with the
baryon-dense nuclear matter case is rather the opposite. The
positive strange-quark chemical potential ($\mu_s > 0$)
corresponds to a colour insulator (hadron gas state), whereas
$\mu_s < 0$ to a colour (super)conductor (deconfined partonic
state). An even simpler system, such as an ideal, three
dimensional Bose gas, has also a well defined chemical potential
as we move above the Bose - Einstein condensation critical
temperature, where $\mu < 0$, or below this critical point, where
$\mu = 0$.

An important argument of this work is that the light and strange
quark fugacities $\lambda_q$ and $\lambda_s$, as well as the
temperature are attributed to the equilibrated primordial state,
to which Eq's (1,12) refer. Only in this case one would expect to
observe negative strange-quark chemical potential and temperature
in excess of those corresponding to deconfinement, for a state
beyond the HG phase. This appears at first as "impossible", since
hadronization always takes place on the
deconfinement-hadronization line, separating the HG phase from the
DQM and, therefore, these quantities should attain values only on
this line. That is, always  $\mu_s = 0$, and for the
sulfur-induced interactions for example, which have $\mu_q \sim
80~MeV$, a temperature $T \sim 176~MeV$ [15, 21].

To overcome this apparent difficulty, we propose without proof
that the conservation of fugacities $\lambda_i~(i=u,d,s)$ is a
characteristic property of strong interactions and thermodynamic
equilibration in general, affecting thermally and chemically
equilibrated states throughout the phase diagram. That is, the
quark fugacities $\lambda_i$, once fixed in a chemically
equilibrated primordial state, are constants of the entire sequent
evolution process. This is contingent on an isentropic expansion
and hadronization via a sudden, fast process without mixed phase.
The relation between $\lambda_i$ and $T$, and hence between
$\mu_i$ and $T$ in the primordial state\footnote[3]{The quantities
$\mu_s, \mu_q$ and $T$ are the Lagrange multipliers of
strangeness, baryon number and energy, attaining their values in
the equilibrated state.} is defined by the strangeness neutrality
equation, obtained from the partition function by imposing
strangeness conservation. The proof of the fugacity conservation
ansatz will require a precise knowledge of the EoS of the DQM
state, so that the variation of the volume, quark density,
temperature, chemical potentials and entropy can be studied from
equilibration to hadronization. This is beyond the scope of the
present paper. It is only stated here in order to give a possible
explanation of an observation of negative strange-quark chemical
potential, which is not a characteristic of the hadronic phase, as
well as of temperatures higher than the maximum temperature for
deconfinement ($T_d \sim 175~MeV$ at $\mu_q \sim 80~MeV$, as given
by SSBM). For the latter, there might be already such an
observation in the $S+S$ interactions at 200 $AGeV$, as discussed
earlier. If proven true, the above statements would have
far-reaching consequences for defining and understanding the
thermodynamic characteristics of the primordial state. They would
suggest that, if nuclear interactions create equilibrated states
beyond the HG phase in the deconfined region with finite net
baryon number density, if the path to hadronization is isentropic
and if the confinement transition is explosive without a mixed
phase, one may determine the thermodynamic quantities of the state
from the strange hadron yields, by employing an appropriate EoS.
If negative $\mu_s$-values, together with temperatures in excess
of $T_d$ are confirmed at the proposed RHIC energies of $20 <
\sqrt{s} < 100~AGeV$, it will be a profound observation. It will
indicate that negative strange-quark chemical potential is indeed
a unique and well-defined signature of deconfinement, identifying
the partonic phase, and that $\mu_s$ is the first experimentally
accessible order parameter of deconfinement. Of equal importance
will be the ability to determine the quantities $\mu_q$, $\mu_s$
and $T$, hence, to calculate accurately the energy density and
entropy of equilibrated primordial states situated beyond the
hadronic phase, in the deconfined-quark region.

On the basis of the present analysis we also conclude that the
$S+S$ collision at 200 $AGeV$ at the SPS may be the first
interaction to have produced an equilibrated partonic state beyond
the hadronic phase. On the other hand, the thermodynamic
parameters of the primordial state of the $Pb+Pb$ interaction at
the highest energy of 158 $AGeV$ strongly indicate that its
location is well within the hadronic phase.

\clearpage
\appendix
\section{Order parameters and the phase diagram}

Since an empirical partition function for the DQM phase is used,
Eq. (12), containing the parameters $R_d(T)$ and $R_{\chi}(T)
\equiv 1 - R_d(T)$, it is important to study the behaviour of the
strange-quark chemical potential in the DQM region for the
different approximations of $R_d(T)$. Figure A.1 is a plot of
$\mu_s$ vs. $T$, obtained for $\mu_q=0.45T$ and the two different
functions for $R_d(T)$. It is clear that the DQM region is
characterized by negative strange-quark chemical potential,
independently of which $R_d(T)$ function is used in the EoS.
Therefore, the change in the sign of the strange-quark chemical
potential, from positive to negative, is a unique and well-defined
indication of the quark-deconfinement phase transition and does
not depend on the phenomenological parameters of the model.

\begin{figure}[htb]
\begin{center}
\epsfbox{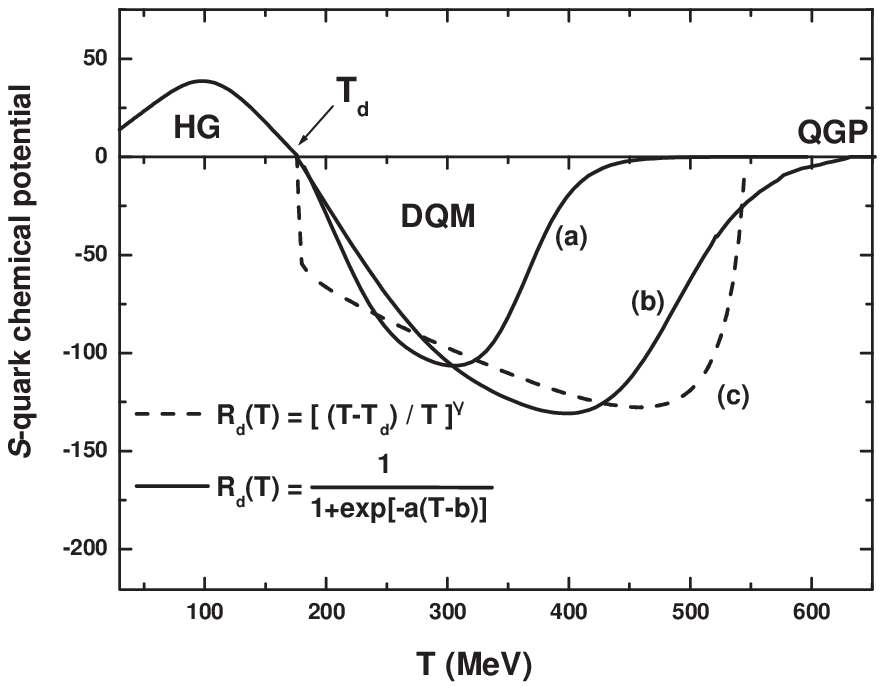}
\end{center}
\caption{Plot of the strange-quark chemical potential as a
function of the temperature and the two different functions for
$R_d(T)$, obtained for (a) {$\alpha=0.05,~\beta=229$}, (b)
{$\alpha=0.03,~\beta=268$} with Eq. (7) and (c) for $\nu=0.1$ with
Eq. (8), respectively.}
\end{figure}

\section{Order parameters and the effective quark masses}

The parameter $R_{\chi}(T)$ determines the effective quark and
hadron masses. The use of temperature dependent quark masses is
one of the essential aspects of the model, since it is a dynamical
term in the equation of state. On the other hand, more precise
theoretical models such as the NJL or the (non)linear sigma model,
have also studied the temperature dependence of the light and
strange quark masses. Figures B.1, B.2 plot the strange and light
quark masses as a function of temperature, using the two different
approximations of the order parameter, whereas Figure B.3 exhibits
the same quantities but within the NJL model. The quark mass
scaling used in our model is consistent with the results of an
effective field theory model, despite the difference for $m_s^*$,
which can be corrected by choosing an appropriate $R_{\chi}(T)$
for strange quarks. This gives strong support to our
approximations and results.
\begin{figure}[htb]
\begin{center}
\epsfxsize=260pt \epsfysize=210pt \epsfbox{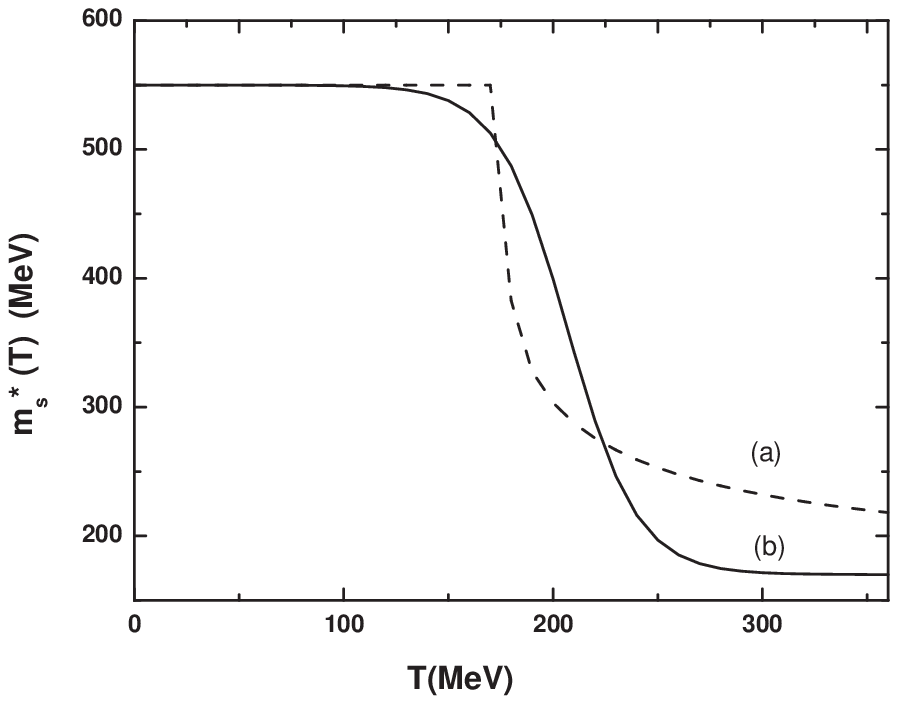}
\end{center}
\caption{The effective quark mass $m_s^*(T)$ calculated using the
expressions of Eq.(8) with $\nu=0.2$, $c=1$ in (a) and Eq.(7) with
$\{\alpha=0.03,~\beta=268\}$ in (b).}
\end{figure}

\begin{figure}[hb]
\begin{center}
\epsfxsize=260pt \epsfysize=210pt \epsfbox{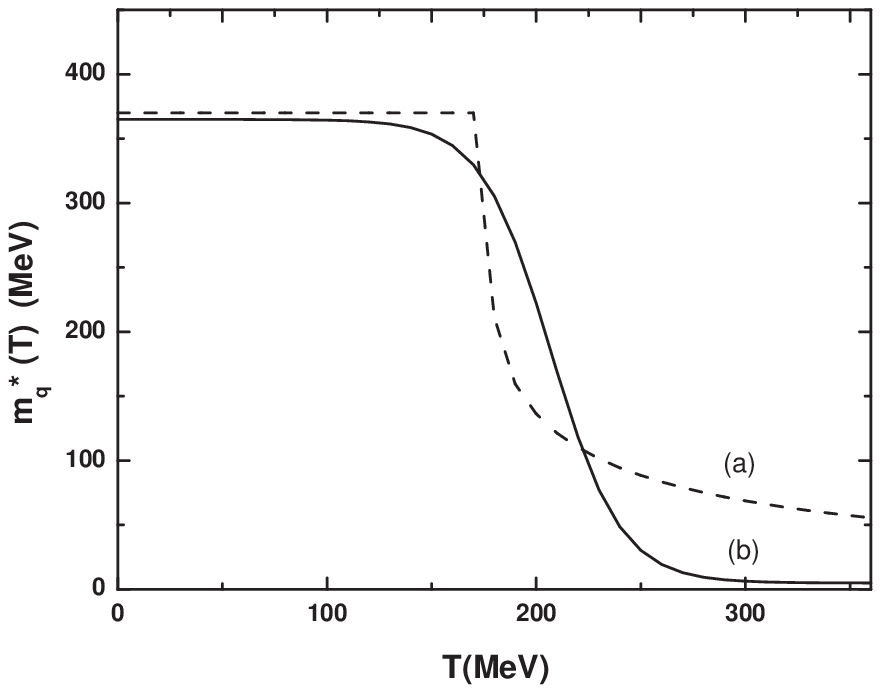}
\end{center}
\caption{The effective quark mass $m_q^*(T)$ calculated using the
expressions of Eq.(8) with $\nu=0.2$, $c=1$ in (a) and Eq.(7) with
$\{\alpha=0.03,~\beta=268\}$ in (b).}
\end{figure}


\newpage
\section*{References}

\begin{thebibliography}{0}
\bibitem {1} Marciano W, Pegels H, {\it Phys. Rep.}
{\bf 36}, 137 (1978)

\bibitem {2} Shuryak E V, {\it Phys. Lett.} {\bf B107}, 103
(1981) \\ Pisarski R D, {\it Phys. Lett.} {\bf B110}, 155 (1982)

\bibitem {3} H. Satz, {\it Nuovo Cim.} {\bf A37}, 141 (1977)

\bibitem {4} Panagiotou A D, Mavromanolakis G, Tzoulis J,
{\it Proc. Int. Conf. "Strangeness in Hadronic Matter"}, 449
(1995)

\bibitem {5} Panagiotou A D, Mavromanolakis G, Tzoulis J,
{\it Strangeness '96 Conf.}, Budapest, {\it Heavy Ion Physics}
{\bf 4}, 347 (1996)

\bibitem {6} Panagiotou A D, Mavromanolakis G, Tzoulis J,
{\it Phys. Rev.} {\bf C53}, 1353 (1996)

\bibitem {7} Karsch F, {\it Nucl. Phys.} {\bf A698}, 199 (2002)

\bibitem {8} Asprouli M N and Panagiotou A D, {\it Phys. Rev.}
{\bf D51}, 1086 (1995)

\bibitem {9} Kapoyannis A S, Ktorides C N and Panagiotou A D,
{\it J. Phys.} {\bf G23}, 921 (1997)

\bibitem {10} Engels J, Kaczmarek O, Karsch F and Laermann E,
{\it Nucl. Phys.} {\bf B558}, 307 (1999)

\bibitem {11} Satz H, {\it Proc. XXIII International
Conference on High Energy Physics}, Berkeley 16-23 July 1986, ed.
S.C. Loken World Scientific Singapore

\bibitem {12} Kajantie K and Kapusta J, {\it Ann. Phys. NY}
{\bf 160}, 477 (1985)

\bibitem {13} Klevansky S P, {\it Rev. Mod. Phys.}
{\bf 64}, 649 (1992)

\bibitem {14} Cabbibo N, Parisi G, {\it Phys. Lett.}
{\bf B59}, 67 (1975)

\bibitem {15} Kapoyannis A S, Ktorides C N, Panagiotou A D,
{\it Phys. Rev.} {\bf C58}, 2879 (1998)

\bibitem {16} Katsas P {\it et al}, {\it work in progress}

\bibitem {17} Ahle L {\it et al} (E-802 Collaboration),
{\it Phys. Rev.} {\bf C57}, 466 (1998); {\it Phys. Rev.} {\bf
C60}, 044904 (1999)

\bibitem {18} Bachler J {\it et al} (NA35 Collaboration),
{\it Z. Phys.} {\bf C58}, 367 (1993)\\ Alber T {\it et al} (NA35
Collaboration), {\it Z. Phys.} {\bf C64}, 195 (1994)

\bibitem {19} Afanasiev S V {\it et al} (NA49 Collaboration),
{\it Phys. Lett.} {\bf B491}, 59 (2000)\\ Afanasiev S V {\it et
al} (NA49 Collaboration), {\it J. Phys.} {\bf G27}, 367 (2001)

\bibitem {20} Kapoyannis A S, Ktorides C N and Panagiotou A D,
{\it Phys. Rev.} {\bf D58}, 034009 (1998)

\bibitem {21} Kapoyannis A S, Ktorides C N and Panagiotou A D,
{\it Eur. Phys. J.} {\bf C14}, 299 (2000)

\bibitem {22} Kapoyannis A S, Ktorides C N and Panagiotou A D,
{\it J. Phys.} {\bf G28}, L47 (2002)

\bibitem {23} Becattini F, {\it J. Phys.} {\bf G23}, 1933 (1997)

\bibitem {24} Sollfrank J, {\it J. Phys.} {\bf G23}, 1903 (1997)

\bibitem {25} Becattini F, Gazdzicki M, Sollfrank J,
{\it Eur. Phys. J.} {\bf C5}, 143 (1998)

\bibitem {26} Becattini F, Cleymans J, Redlich K,
{\it Phys. Rev.} {\bf C64}, 024901 (2001)

\bibitem {27} Cleymans J, Redlich K,
{\it Phys. Rev.} {\bf C60}, 054908 (1999)

\bibitem {28} A. D. Panagiotou, P. G. Katsas, E. Gerodimou
{\it J. Phys.} {\bf G28}, 2079 (2002)

\bibitem {29} Hagedorn R, {\it Riv. Nuov. Cim.} {\bf 6}, 1 (1983) ;
{\it Proc. Advanced NATO workshop: Hot Hadronic Matter. Theory
$\&$ Experiment}, Divonne-les-Bains July 1994, ref's therein.

\bibitem {30} Braun-Munzinger P, Magestro D, Redlich K and Stachel
J, {\it Phys. Lett.} {\bf B518}, 41 (2001)

\bibitem {31} Moulopoulos K and Ashcroft N W,
{\it Phys. Rev.} {\bf B42}, 7855 (1990)

\bibitem {32} Stock R, {\it Proc. 18th Winter Workshop on Nuclear Dynamics}, 001-002
(2002) [arXiv:hep-ph/0204032]
\end{thebibliography}
\end{document}